\documentclass[12pt]{iopart}
\usepackage{setstack}
\usepackage{graphicx} 
\usepackage{bm}
\usepackage{iopams}
\renewcommand{\contentsname}{}
\usepackage[usenames]{color}

\begin{document}
\title{Fourier microscopy of single plasmonic scatterers}
\author{Ivana
Sersic, Christelle Tuambilangana and A Femius Koenderink}
\address{Center for Nanophotonics, FOM Institute for Atomic and
Molecular Physics (AMOLF), Science Park 104, 1098 XG Amsterdam, The
Netherlands}
\ead{i.sersic@amolf.nl}
\begin{abstract}
We report a new experimental technique for quantifying the angular distribution of light scattered by single plasmonic and metamaterial nanoscatterers, based on Fourier microscopy in a dark field confocal set up. This new set up is a necessary tool for quantifying the scattering properties of single plasmonic and meatamaterial building blocks, as well as small coupled clusters of such building blocks, which are expected to be the main ingredients of nano-antennas, light harvesting structures and transformation optics. We present a set of measurements on Au nanowires of different lengths and show how the radiation pattern of single Au nanowires evolve with wire length and as a function of driving polarization and wave vector.
\end{abstract}
\pacs{42.25.-p, 42.30.Kq}
\date{May 16, 2011}
\submitto{New. J. Phys., May 16th, 2011}
\maketitle

\section*{Contents}
\tableofcontents

\section{Introduction}
The main focus in the fields of plasmonics and metamaterials are electro-magnetic resonances that occur in sub-wavelength building blocks ~\cite{Pendry01, Soukoulis07a, Shalaev07, Lederer07, Soukoulis07b}. Plasmonic and metamaterial resonances play an important role in nanoscale quantum optics ~\cite{AkimovChang,Curto10}, enhanced sensing ~\cite{VanDuyne08}, solar cells ~\cite{Polman10}, and transformation optics ~\cite{Pendry10}. In order to build functional devices and materials from plasmonic and metamaterial scatterers, it is important to quantify how single building blocks scatter ~\cite{vdHulst, Abajo07, Husnik08, Sersic09, Pors10, Sersic11}. Such a quantification requires to not only determine the magnitude of their scattering, absorption and extinction cross section ~\cite{Husnik08}, but also to measure the distribution of light scattered into each direction ~\cite{Curto10, Shegai11}. Measurements of angular distributions of light are highly challenging as signal levels from single nano-objects are low. Therefore, one often resorts to studying arrays of identical nano-objects. However, in scattering experiments on arrays one typically faces the problem that the radiation patterns of collections of scatterers are dominated by grating diffraction orders (periodic arrays), or speckle (random arrays). Separating this `structure factor' from the angle-dependent radiation pattern of single objects in such measurements is challenging since it is difficult to create a featureless structure factor.
Recently, Fourier microscopy, or `backaperture imaging', has been utilized by several groups as a method to measure angle-dependent radiation patterns. This method is based on collecting the light radiated by a single nanostructure using a standard high NA objective, and the realization that the objective back aperture  contains $k$-space information of the electromagnetic field. It was first applied to measure the radiation pattern of single molecule emitters,  thereby allowing to determine their orientation ~\cite{Novotny04, Patra05}. This method was also used recently ~\cite{Curto10, Wenger11} to measure the directivity of emission for molecules coupled to a Yagi-Uda antenna and plasmon nano-apertures. 

Surprisingly, the application of Fourier microscopy to scattering experiments is much less widespread ~\cite{LeThomas07, Drezet08, Randhawa10}.
For the geometry of metal particles on transparent substrates  that is possibly of biggest interest in the fields of plasmonics and metamaterials, Fourier microscopy is most difficult to implement due to issues with background light. The only implementation that we are aware of ~\cite{Huang08} is limited to a very small subset of collection angles, namely only those angles above the total internal reflection angle in glass.
Shegai \etal ~\cite{Shegai11} have applied back aperture imaging to study radiation from the endpoints of nanowires. However, in their work light can only be collected from parts of a bigger structure, as the focused illumination applied to one part of a structure to  excite it needs to be removed by spatial filtering. So far, however, the possibility of recording angle resolved scattering data over a full objective NA for single nano-objects on simple transparent substrates has not been reported.  In this paper we present an optical set up that can record radiation patterns for such subwavelength objects upon excitation with a well defined incident wave vector and polarization. In this way, it is possible to for the first time quantify how, e.g., single elements in plasmon antennas and metamaterials scatter.
We present a set of measurements on radiation patterns of Au nanowires of different lengths. We demonstrate that while ultrashort Au nanobars radiate as a single point dipole, longer single nanowires can be understood as a collective oscillation of a line array of  dipoles.

\section{Experimental setup}
\begin{figure}
\begin{center}
\includegraphics[width=9 cm]{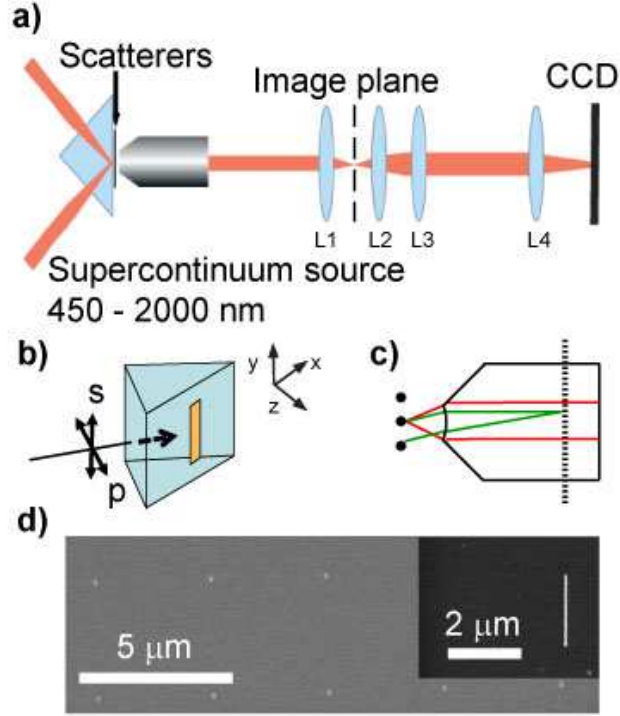}
\caption{(a) A schematic of the Fourier microscope set up. The set up consist of a high NA objective (NA=0.95), a set of telescope lenses L1 and L2 with equal focal lengths ($f$=50 mm), a Fourier lens L3 ($f$=200 mm) and a tube lens L4 ($f$=200 mm) that focuses the image on a silicon Charge Couple Device (CCD). To select a single scatterer, we place a pinhole in the image plane between L1 and L2.
(b) A schematic representing the position of a nanowire on the prism front facet with a long axis oriented parallel to the $y$-axis, excited with $s$ or $p$ polarized incident light. (c) Ray diagram demonstrating the front and back aperture image planes. The light scattered by nano-objects is collected by the objective which forms a collimated beam (red lines). Each point in the back aperture of the objective (dashed black line) corresponds to a different wave vector (green lines) scattered by a nano-object. (d) SEM image of an array of 200 nm long, 50 nm wide and 30 nm thick nanobars arranged in a periodic lattice with 4 $\mu$m lattice spacing. The inset shows an SEM zoom in of a 2 $\mu$m long, 50 nm wide and 30 nm thick Au nanowire. }\label{setup}
\end{center}
\end{figure}

Our homebuilt experimental set up is sketched in figure~\ref{setup} (a). A basic design constraint that we impose is that the driving field has a well defined wave vector and polarization. Therefore, we opt for an input beam with a large focus, i.e. 30~$\mu$m spot size, that has $|\mathbf{\Delta k}|/|\mathbf{k}| \leq 0.02$. Since any typical plasmonic scatterer has a cross section that is at most 10 times its geometrical cross section (~0.1~$\mu$m$^{2}$), just 10$^{-4}$ of the incident power is scattered in total per object. In a Fourier microscope this is not all collected by a single detector (as in an imaging microscope), but is spread over a detector array with ~10$^4$ channels. Hence, each angular detection channels receives only about one photon per 10$^8$  incident photons, making a very bright source and excellent background suppression in the set up a necessity.
Therefore, we have used a supercontinuum light source (Fianium) in combination with an acousto-optical tunable filter (AOTF) for frequency selection. The combination of Fianium and AOTF provides at least 0.1 mW in a 5 nm bandpass windows centered at any wavelength in the visible or near-infrared.
Since scattering does not differ in frequency from the input beam, as in fluorescence experiments, it is not possible to differentiate the signal from the background signal by means of frequency filtering. Therefore, we have utilized dark field microscopy in total internal reflection (TIR) mode to excite our structures.

The samples are placed on the front facet of a glass prism with index matching liquid between the glass substrate and the prism ($n$=1.45), as schematically depicted in figure~\ref{setup} (b). The incident beam impinges on the sample at an incidence angle $\theta_I$ that is greater than the critical angle $\theta_C$ needed for TIR. Our structures are hence excited by an evanescent wave that has a wave vector component along the prism equal
to the parallel wave vector of the incoming light ($\mathbf{k}_{||}$),
and with some evanescent decay from the interface, set by $k_z=\sqrt{|\mathbf{k}_{||}|^2-(\omega/c)^2}$. The scattered light is collected by a high NA 60x Olympus objective (NA=0.95), meaning that we collect scattered parallel wave vectors in the range
$-0.95\frac{\omega}{c}\leq\mathbf{k}_{||}\leq0.95\frac{\omega}{c}$. The microscope objective is mounted on a Newport ultralign micrometer stage for fine focusing and positioning, which is set at the end of an optical rail.
This rail furthermore contains a set of two telescope lenses in order to create an intermediate real space image plane. We use this plane to select light
from just one single nano-object, by placing a pinhole mounted on a flip mount. In our setup, we use a telescope lens with $f$ = 50 mm, which implies a magnification of 27x.  Therefore one could use commercially available 50 to 120~$\mu$m pinholes for spatial selection.
In general, care must be taken that pinholes are small enough to exclude adjacent objects, yet large enough that Airy diffraction rings from the pinhole do not dominate the Fourier image. Moreover, we have found it necessary to deal with the fact that commercial pinholes are not circular, and the fact that the abrupt edges of binary pinholes always diffract.
Therefore,  we use Gaussian graded pinholes defined by  digitally transferring  16-bit black and white tiff-file definitions of gaussian circular grayscale patterns in photographic black and white slide film by laser writing (www.colorslide.com).  These pinholes have a smooth, apodized transmission in Fourier space. In order to block residual transmission through the nominally non-transmitting part of the film away from the pinhole (for which we estimate  an optical density of 3), we glue the apodized pinholes onto a stainless steel thin film with a mechanically drilled 300~$\mu$m hole. When used as a simple  imaging microscope, the  set up contains an additional of a tube lens ($f$=200 mm)  that images the pinhole plane on a CoolSnap EZ Silicon CCD camera.

In order to  retrieve the radiation pattern of a single nano-object, i.e. the $k$-space image, we place a flippable Fourier lens between telescope and tube lens,  at a distance $4f_{\mathrm{telescope}}+f_{\mathrm{Fourier}}$ from the back focal plane of the objective to focus the CCD camera on the microscope back aperture, rather than on the sample plane (i.e., at infinity in our infinity corrected microscope). Figure~\ref{setup} (c) shows a ray tracing image of the light collected by the objective in real (red lines) and $k$-space (green lines). Independent of the position of the object, each radiation angle is focused onto a unique location in the back aperture of the objective (dashed black line).
\begin{figure}
\begin{center}
\includegraphics[width=9 cm]{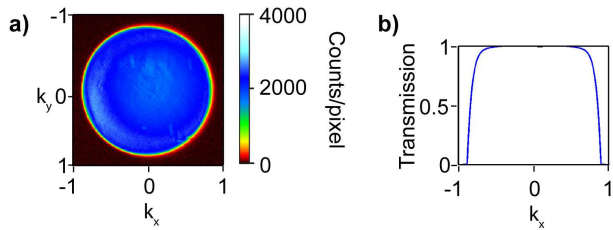}
\caption{(a) Fourier space image of a uniform layer of 100 nm dye-doped fluorescent beads excited with unpolarized incident light at 480 nm. The total exposure time is 0.4 s. (b) The blue line is the transmission function extracted from the measurements in (a).}\label{transmission}
\end{center}
\end{figure}

We calibrated the objective transmission function as a function of collection angle by measuring the intensity distribution in the back aperture from fluorescence emitted by dye-doped fluorescent beads (Invitrogen Fluospheres F8800) emission ~\cite{Dai05}. We have deposited a uniform layer of 100 nm beads by spincoating a 5$\%$ concentrated solution. The dye molecules have an emission peak at ~535 nm. A set of band pass filters was used in the incoming beam (tuned to 480 nm) and the detection beam in order to selectively collect only the fluorescent light coming from the layer of dye molecules. The isotropic emission collected by our objective is shown in figure~\ref{transmission} (a). It has a non-uniform intensity distribution in Fourier space firstly because equidistant angles are not equidistant in the $\mathbf{k}_{||}$ space that our CCD images, and secondly because of the angle dependent apodization function $T(k_{||})$ of the objective. For an angularly isotropic emission we expect the collected intensity to vary as
\begin{equation}
I(k_{||})=\frac{P}{\cos(\mbox{arcsin}(k_{||}))}T(k_{||}), \label {apodization}
\end{equation}
where $k_{||}={|\mathbf{k}_{||}|}/{k_0}$. In our measurements, $T(k_{||})$ is angle independent  for angles between $0\leq |\sin\theta|\leq 0.6$, rolling off to  50\% of its peak value at $\sin\theta=0.86$, as shown in figure~\ref{transmission} (b).  Since the objective is specified for near infrared applications, we expect that the transmission edge, which  essentially reaches zero already at an  NA of 0.89, moves to larger angles for longer wavelengths, where the NA is specificied as 0.95 by the objective manufacturer ~\cite{Dai05}.
The relation between pixel on the CCD camera and wave vector emitted in the object plane is easily calibrated by using the Fourier microscope
without spatial filter. Since we use periodic structures, the collected pattern consists of grating diffraction orders that are equidistant in $k_{||}$-space.  Indeed, in accord with the Abbe sine condition by which the objective is designed, we retrieve equidistant lattices of dots in $k_{||}$-space  on the CCD camera, the spacing of which serves as calibration. In our set up, the full NA of the objective corresponds to a diameter of approximately 300 pixels on the camera.

\section{Results and discussion}
\subsection{Fourier microscopy of an array of ultrashort gold nanobars}
In order to demonstrate the potential of Fourier microscopy of single nano-objects, we have fabricated samples with Au nanobars of different lengths. The shortest bars that we fabricated (200 nm long, 50 nm wide, 30 nm high) are sub-wavelength (figure~\ref{setup}), and hence expected to have only weakly directional scattering patterns. In contrast, the longest bars  (4~$\mu$m long, 50 nm wide, 30 nm high) are so long that they are expected to have several plasmon guided mode Fabry-Perot resonances and potentially quite directional scattering patterns ~\cite{Ditlbacher05,Vesseur07,Dorfmuller11}. The inset of figure~\ref{setup} (d) shows a scanning electron micrograph (SEM) of a single 2 $\mu$m long and 50 nm thick nanowire. The nanobars were fabricated by defining lines in electron beam lithography in ZEP-520 resist, employing thermal evaporation of Au, and subsequently lift-off in 1-methyl-2-pyrrolidinone (NMP). The nanowires are arranged in periodic square arrays with lattice spacing equal to 20 times the nanowire length, as shown in figure~\ref{setup} (d).
\begin{figure}
\begin{center}
\includegraphics[width=9 cm]{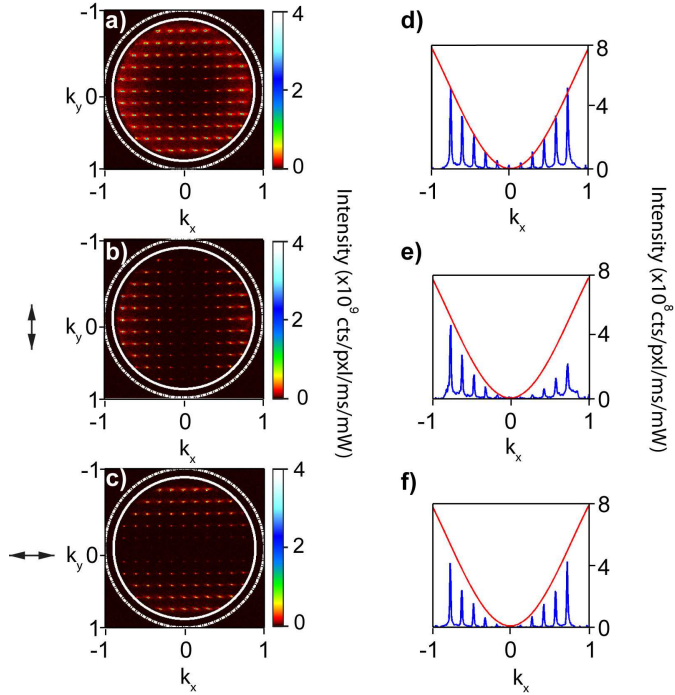}
\caption{(a) Fourier space image of a periodic square array of 200 nm long, 50 nm wide and 30 nm thick Au nanobars with 4 $\mu$m lattice spacing excited with a $p$-polarized incident light at 600 nm with ~20 $\mu$W. The total exposure time is 10 ms. (b), (c) Radiation patterns under the same illumination conditions as in (a) and polarization analyzed along (b) and across (c) the nanobar length, as denoted by the black arrows. (d), (e), (f) Fitted sin$^2$ (red line) to cross sections along the central column ((a), (b)) and row ((c)) of grating orders in our measurements in (a), (b) and (c), respectively, with a scaling factor of 11$\times10^8$ and an offset of 3$\times10^6$ counts per pixel, per mW incident power and per ms exposure time.}\label{grating}
\end{center}
\end{figure}
The pitch is large enough to avoid any coupling between objects, yet small enough so that we can easily find fields of objects in widefield or darkfield microscopy. Figure~\ref{grating} (a), (b), and (c) show Fourier images of a periodic array of 200 nm Au nanowires with 4 $\mu$m lattice spacing, without spatial filtering, excited by $p$-polarized light (a) and polarization analyzed along [(b)] and across [(c)] the nanobar length. The white edge around the intensity distribution depicts the measured NA of our objective (full white line), corresponding to all wave vectors up to 89 $\%$ of the maximum wave vector in free space (dashed white line), while the center of the image corresponds to $k_{||}$=0. The Fourier space is clearly dominated by grating orders, as expected for a periodic array.

Interestingly, not all grating diffraction orders are equally intense. We expect ~\cite{Abajo07} that the Fourier space scattering of a periodic array of scatterers is the product of the radiation pattern of each single scatterer and the structure factor of the array that is  a set of $\delta$-peaks at the vectors $\mathbf{k}_{||} + \mathbf{G}$ (where $\mathbf{G}$ is any of the reciprocal lattice vectors).  In other words, the radiation pattern of an array makes up a sparse sampling of the single object radiation pattern.  In this data we recognize that the central orders near $k_{||}=0$ are much weaker than the orders at larger angles, as seen from cross sections through our data along $k_x$ in figure~\ref{grating} (d), (e), and (f) (blue lines).
For this data set, we have used excitation with $p$-polarized light, obtained by placing a polarizer  in the incident beam. For $p$-polarized driving we expect that each nanobar obtains a large, out of plane dipole moment.   For out-of-plane dipoles we expect that the single building block radiation pattern is strong at large angles and weak near $k_{||}=0$, since the radiation pattern of a single dipole is given by $P \sim\sin^2 \theta = k_{||}^2$. We demonstrate this behaviour by fitting the sin$^2$ to the cross sections in figure~\ref{grating} (d), (e) and (f) (red lines). The fact that the grating diffraction orders represent a discrete sampling of this single block radiation pattern is further confirmed by  a  polarization analysis, realized by placing a second polarizer immediately after the microscope objective. The grating diffraction orders  reveal a  radial polarization around the intensity node at $k_{||}=0$, consistent with the radial polarization expected for a single out-of-plane dipole moment.
\begin{figure}
\begin{center}
\includegraphics[width=9 cm]{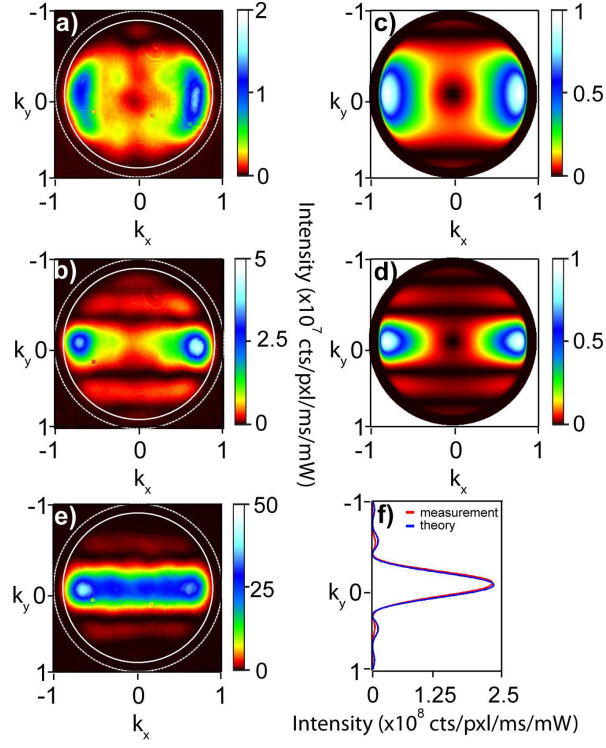}
\caption{(a) Fourier space image of a 1 $\mu$m long, 50 nm wide and 30 nm thick Au nanowire excited by  $p$-polarized incident light. (b) Fourier space image of a 2 $\mu$m long, 50 nm wide and 30 nm thick Au nanowire excited by a $p$-polarized incident light. (c) and (d) are calculated radiation patterns of 2 $\mu$m and 1 $\mu$m long Au nanowires multiplied by the transmission function, respectively.
(e) Fourier space image of a 2 $\mu$m long, 50 nm wide and 30 nm thick Au nanobar excited with  $s$-polarized incident light. The nanobars are excited at 725 nm and ~200 $\mu$W. The total exposure time is 1 s in (a), 100 ms in (b), and 10 ms in (e). (f) An average cross-cut of (e) along $k_y$ (red curve) agrees well with the calculated sinc$^2$ behavior (blue line).}\label{vertical} 
\end{center}
\end{figure}
\subsection {Fourier microscopy of single scatterers}
The measurements in figure~\ref{grating} (a) and (b) show that Fourier microscopy of arrays of nano-objects is strongly limited by the fact that angle-dependent scattering strength is sampled only at a sparse set of points, set by the grating
diffraction orders. We will now present measurements that show that it is possible to even measure radiation patterns of single sub-wavelength plasmonic scatterers with our Fourier microscope. To this end we flip the spatial selection filter into the image plane between lens L1 and L2 in the telescope (figure~\ref{setup}), to select a single scatterer.
When the spatial selection filter is in place, the grating orders
disappear and we can observe the full structure of the radiation pattern.  In figure~\ref{vertical} (a) and (b) we show radiation patterns (measured without a polarizer in the collection path) of 2~$\mu$m and 1~$\mu$m Au  bars that are perpendicular to the scattering plane of the incident, $p$-polarized light.  In this configuration, the incident beam excites the entire object in phase ($k_{||}=0$), since the extent of the nanowires as measured along the incident wave vector is only 50 nm (the wire width). The images reveal low to no intensities in Fourier space around $k_{||}$=0, with most of the intensity concentrated at high angles, specifically
at high $k_x$, yet small $k_y$.
Furthermore, upon comparison of the radiation pattern of the 2~$\mu$m and 1~$\mu$m wire, we observe that while both radiation patterns are confined to a narrow range of wave vectors near $k_y=0$, this confinement is roughtly two times greater for the longer wire. The measurements futher show secondary maxima around the main lobe centered at $k_y$=0. An increasing concentration of scattered radiation around a central direction with wire length was also noted by Shegai \etal ~\cite{Shegai11}, demonstrating stronger directionality for wires with a higher geometrical aspect ratio. Complementary measurements for $s$-polarized driving on a 2$~\mu$m Au bar (figure~\ref{vertical} (e))  also show strong directionality when dipole moments are excited along the bar. Again the radiation pattern is confined to a narrow region around $k_y=0$. A striking difference with $p$-polarized driving, however, is that intensity is more uniformly distributed along the $k_x$-axis, with no apparent reduction of intensity at $k_{||}=0$. This set of measurements clearly shows the main advantage of our Fourier microscope, i.e., the ability to map the full back aperture of our objective for light scattered by a single nano-object. For instance, the striking difference in radiation pattern depending on polarization of the driving would not have been noticed in set ups ~\cite{Curto10, Shegai11, Huang08} that collect only large wave vectors beyond total internal reflection.

In order to understand the radiation patterns of our nanowires, we implement a simple model. We hypothesize that the radiation pattern of wires can be simply described as that of a set of equivalent point dipoles arranged in a line over the length of the wire.  For this case, we expect that different volume elements along the length $L$ of the wire are all excited in phase, and with equal incident amplitude and are all polarized along the optical axis of the setup. Due to the slight phase differences accumulated for waves travelling to a given observation point in the far field  from different positions on the wire, the radiation pattern of a line of dipoles is modified by the form factor of the wire, obtained by integrating over the  wire.
In the theory of microscopic imaging with high NA aplanatic lenses that satisfy the Abbe sine rule, it is well known that the back aperture field can be found directly from the field on a reference sphere of radius $f$, where $f$ is the objective focal distance ~\cite{Torok98,Haeberle03}. If we divide the wire into area elements $\mathrm{d}x\mathrm{d}y$, the field on the reference sphere will be
\begin{equation}
\mathbf{E}(\theta,\phi)\sim\frac{\mathbf{E}_{\mathrm{dip}}^{\mathrm{far}}(\theta,\phi)}{f}\int_{\mbox{wire area}}\,e^{-i k_0 R_f(\theta,\phi,x,y)}\mathrm{d}x\mathrm{d}y\label{Eq:general}
\end{equation}
where ${\mathbf{E}_{\mathrm{dip}}^{\mathrm{far}}(\theta,\phi)}/{f}$ is the electric field amplitude given by the radiation pattern of a single dipole.
For $f \gg L$, the distance from radiator to reference sphere ${R}_f(\theta,\phi,x,y)$ simplifies to ${R}_f=f-\frac{\mathbf{k}_{||}}{k_0}\cdot\mathbf{r}_{||}$, where $\mathbf{k}_{||}=k_0(\cos\phi\sin\theta,\sin\phi\sin\theta,\cos\theta)$, and where $\mathbf{r}_{||}=(x,y,0)$ . For a wire of length $L$ oriented along the $y$-axis, and of infinitesimally small width,  the integral simplifies to
\begin{equation}
\mathbf{E}(\mathbf{k}_{||})=\mathbf{E}_{\mathrm{dip}}^{\mathrm{far}}(\mathbf{k}_{||})\frac{e^{i k_0f}}{f}\int\limits_{-L/2}^{L/2}\,e^{-ik_y y} \mathrm{d}y
 \propto
 \mathbf{E}_{\mathrm{dip}}^{\mathrm{far}}(\mathbf{k}_{||})\mathrm{sinc}(k^y_{||}L/2),
\label{Eq:sinc}
\end{equation}
\begin{figure}
\begin{center}
\includegraphics[width=9 cm]{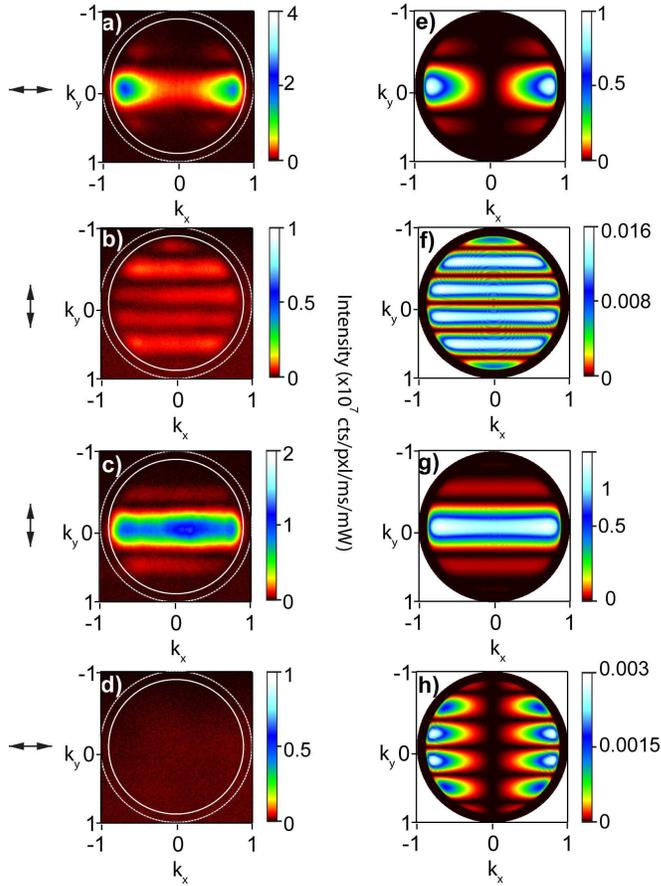}
\caption{(a), (b) Polarization analyzed radiation patterns of 2 $\mu$m long, 50 nm wide and 30 nm thick Au nanowires excited by a $p$-polarized incident light at 650 nm with ~200 $\mu$W along (a) and across (b) the nanowire length, as denoted by the black arrows. The exposure time in (a) is 300 ms and in (b) 1 s. (c), (d) Polarization analyzed radiation patterns for the same nanowire excited by an $s$-polarized incident light at 650 nm and ~200 $\mu$W along (c) and across (d) the nanowire length, as denoted by the black arrows. The exposure time in (c) is 500 ms and in (d) 1 s. (e)-(h) Calculated radiation patterns corresponding to the illumination conditions measured in (a)-(d), respectively, multiplied by the transmission function.}\label{analysis}
\end{center}
\end{figure}
This calculation is analogous to calculating Fraunhofer  diffraction of a slit, but now is applied to nanoscale scatterers with wide radiation patterns that are far from paraxial.  Our calculation predicts that the polarization content is directly inherited from the radiation pattern of a single point dipole multiplied by a sinc function that  applies irrespective of incident or collected polarization.  
The calculation in essence predicts that the radiation pattern of a nanowire is that of a single dipole multiplied by a sinc$^2$ function that ensures that the longer the wire is, the stronger the radiation is confined to the plane transverse to the wire.  The excellent correspondence between the data in figure~\ref{vertical} and the calculation not only concerns the width of the central lobe, set by the {sinc}$^2$ function, but also the appearance of a minimum in scattered intensity at the center of the pattern that is proportional to
$| \mathbf{E}_{\mathrm{dip}}^{\mathrm{far}}(\mathbf{k}_{||})|^2$ . In addition, the calculation also correctly predicts the location of side lobes in the $k_y$ direction. For completeness we  provide a  cross-cut through data and theory, obtained by integrating data in figure~\ref{vertical}(e) along $k_x$. Data (red line) and theory (blue line) in figure~\ref{vertical} (f) are in excellent agreement. The absence of a hole in the radiation pattern in  figure~\ref{vertical}(e) is furthermore consistent with the fact that $|\mathbf{E}_{\mathrm{dip}}^{\mathrm{far}}(\mathbf{k}_{||})|^2$  has no central minimum for in-plane oriented dipole moments.

As noted above, our calculation predicts that the polarization content of the scattered  light  is directly inherited from the radiation pattern of a single point dipole.  To verify this prediction, we place a polarizer directly behind the microscope objective to analyze the polarization of the scattered light, as reported in figure~\ref{analysis} for $p$-polarized excitation. Since the single-dipole riadation pattern is radially  polarized, analyzing the polarization of the 2 $\mu$m nanowire radiation pattern along $k_x$ only retrieves the main lobe of the sinc$^2$ function. For cross polarization, i.e., polarization along the wire, the field $\mathbf{E}_y(\mathbf{k}_{||})$ has a node at $k_y$=0. As a consequence the main lobe is crossed by a nodal line and is strongly suppressed so that it becomes comparable in brightness to the low intensity side lobes at larger $k_y$. Our measurements with the polarization analyzer (figure~\ref{analysis} (a) and (b)) show very good agreement with the calculations (figure~\ref{analysis} (e) and (f)).
We have also polarization analyzed Au nanowires that are perpendicular to the scattering plane, but with $s$-polarized incident light. Figure~\ref{analysis} (c) and (d) show measured radiation patterns that are polarization analyzed along (c) and across (d) the nanowire length. Again, the measurements show very good agreement with the calculations of the same polarization analysis shown in figure~\ref{analysis} (g) and (h). Note that the calculation in figure~\ref{analysis} (g) has much lower intensities than (e) and (f).

\subsection{Radiation pattern of single Au nanowires oriented in the scattering plane}
So far we have considered radiation patterns of single Au nanowires perpendicular to the scattering plane of the incident beam, in which case the whole wire is excited in phase. Many excitations in small nano-objects will benefit from excitation with $\mathbf{k_{||}}$ different from 0. For instance, for understanding the excitations of 1D objects like wires and particle chains,  it would be advantageous to phase-match the excitation wave vector to that of guided modes. Therefore, we have also studied radiation patterns of Au nanowires that are oriented in the scattering plane of the incident wave.
\begin{figure}
\begin{center}
\includegraphics{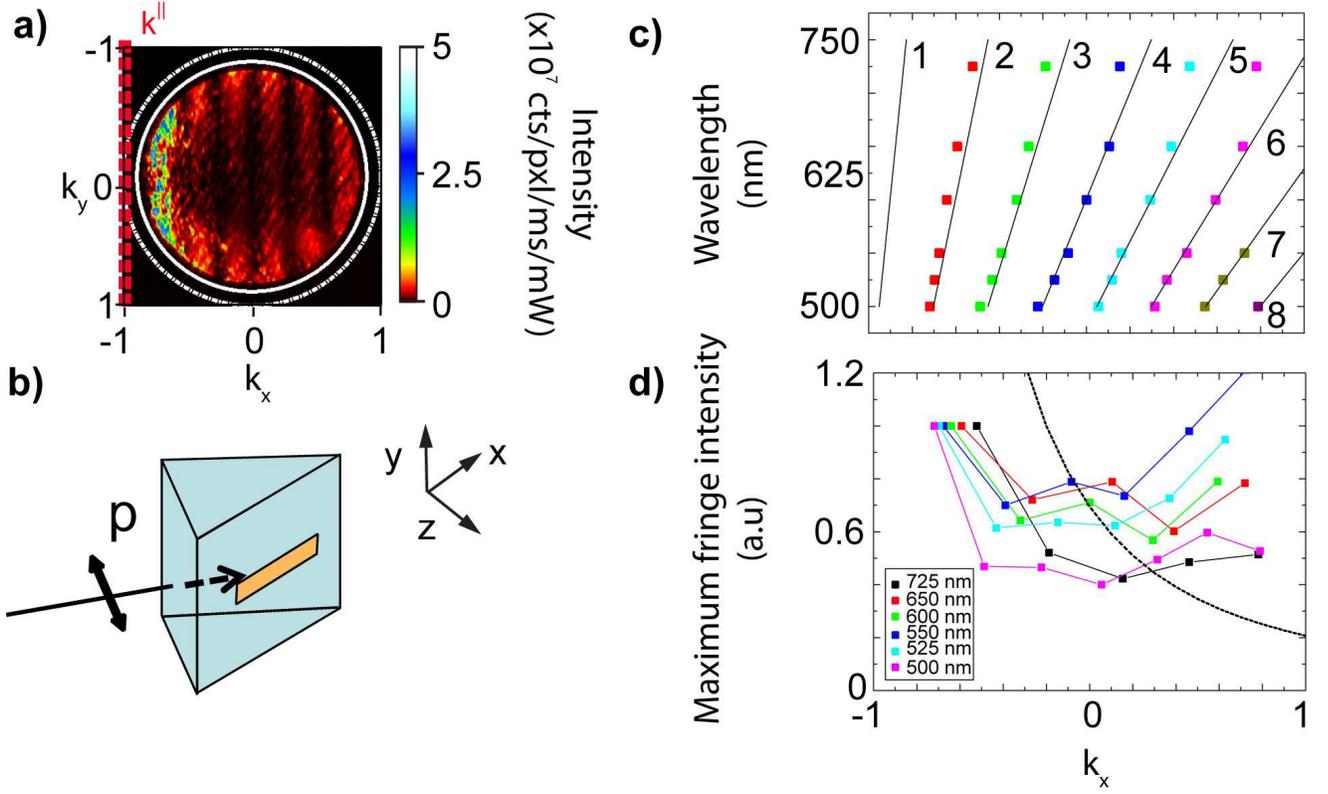}
\caption{ (a) A schematic representing the position of a nanowire on the prism front facet with a long axis parallel to the $x$-axis, excited with $p$ polarized incident light. (b) Fourier space image of a horizontal 2 $\mu$m long, 50 nm wide and 30 nm thick Au nanobar excited with $p$-polarized incident light at 550 nm and 8 $\mu$W. The exposure time is 1 s. (c) Location of
the fringe intensity maxima measured in (b) along $k_x$. The black lines show the predicted position of the intensity lobes away from the zero-order lobe. Symbols are black, red, green, blue, cyan, magenta, dark yellow and purple for fringe orders m=1-8, respectively, marked by numbers. (d) Symbols (connected by lines for clarity): maximum fringe  intensity normalized to the m=2 fringe intensity versus parallel wave vector. The black dashed line depicts the expected $1/|k_{x}^{in}-k_x|^2$ behaviour of the intensity lobes. Fringes with the same wavelength but different orders are color coded. }\label{horizontal}
\end{center}
\end{figure}
Figure~\ref{horizontal} (a) shows a measured radiation pattern of an array of 2 $\mu$m long Au nanowires with 40 $\mu$m lattice spacing. A nanowire that is excited with $p$-polarized light has a radiation pattern dominated by the sinc$^2$ function that is given by the total length of the object, as previously discussed.
However,  now the sinc function is rotated by 90 $^{\circ}$ together with the wire (figure~\ref{horizontal} (b)), and displaced from wave vector $k_{||}=0$ to be centered at the incident wave vector $\bf{k_{||}}$.
Of course, since we are working in TIR, the incident wave vector and hence the main lobe of the sinc$^2$ function is just outside the part of Fourier space accessible to our objective. In our configuration the central lobe is located at the left hand side just outside the Fourier image (red dashed line in figure~\ref{horizontal} (b)). 
A striking difference with measurements with in-phase excitation (figure~\ref{vertical}) is furthermore that the overall collected signal per bar is much weaker. We attribute this weak signal to the fact that the main lobe of the sinc$^2$ function is beyond the light line in air, reducing scattering into the collection side of the set up. It is for this reason that figure~\ref{horizontal} shows data obtained without pinhole on a sample with a very dilute set of wires (40 $\mu$m lattice spacing), rather than with pinhole. Compared to data with pinhole (not shown), the advantage is a large boost in signal, though at the price of obtaining only a sparse sampling of the radiation pattern, due to grating diffraction as in figure~\ref{grating}.

We have measured radiation patterns for $p$-polarized incident light and nanowire orientation for a wide range of incident frequencies. Figure~\ref{horizontal} (b) shows the position of radiation pattern fringe maxima for a 2 $\mu$m nanowire as a function of incident frequency. As expected, the fringe maxima are equidistantly spaced by $k_x/k_0=\lambda/L$, and originate from a central zero-order lobe just outside the diagram. Further, we would expect the fringe intensity to drop off as $1/|k_{x}^{in}-k_x|^2$. However, we don't observe such a monotonic fringe intensity drop off in all our data sets.  Specifically, figure~\ref{horizontal} (c) shows that all fringes are more or less comparable in brightness for $\lambda<$ 650 nm across the whole back aperture. Only for $\lambda=$ 725 nm and above (not shown) do we find a drop off in fringe brightness commensurate with the expected sinc$^2$ tail. A possible explanation is that plasmonic resonances of the wire modify the radiation patterns. Indeed, it has been predicted ~\cite{Dorfmuller11, Taminiau11} that the current distribution excited in a metal wire not only has a component directly proportional  to the incident field (wave vector $k_{||}$), but also due to standing plasmon wave oscillations along the wire. Such standing waves would add extra contributions to the radiation pattern that are again of the form of sinc$(kL/2)$, but centered at $k_x=\pm k_{SPP}$, i.e., at the guided plasmon wave vector. A detailed analysis of fringe intensity versus $k_x$ would allow to extract the dispersion relation of the nanowire. However, in our data, this analysis  is obscured by the fact that plasmon resonances in lithographically fabricated Au wires are generally not very strong due to losses, and by the fact that in total internal reflection illumination, the plasmonic $x$-oriented  mode is only weakly driven. Indeed, for incident angles higher than the critical angle required for TIR, $\theta_I>\theta_C$, the evanescent wave at the surface of the prism has an electric field component along $z$, which for the incidence angle in our experiments ($52^0$)  is  $2\frac{1}{2}$ times greater than the $x$-oriented field. Separating the non-resonant, but strongly excited polarization perpendicular to the wire from the weakly excited but possibly resonant wire plasmon radiation patterns is outside the scope of this paper.   For application of Fourier microscopy to complex resonant structures in general, it is a major challenge to simultaneously control the required incident phase gradient over the structure and achieve the desired polarization, while also remaining in total internal reflection or dark field excitation mode. We suggest that combining Fourier microscopy with wavefront phase shaping ~\cite{Moskvellekoop}\ may be a  promising route to extract further quantitative information to benchmark models such as those proposed in ~\cite{Dorfmuller11,Taminiau11}.

\section{Summary and conclusion}
In conclusion, we have built a Fourier microscope that is suited for measuring the  radiation pattern of single  plasmonic and metamaterial scatterers.
We have successfully measured radiation patterns of single Au nanowires with different lengths down to 200 nm, even though signal levels drastically reduces with size.  Since Fourier microscopes always operate in dark field mode, the incident excitation field is  limited in  polarization and wavefront. We have shown that even with the class of driving fields available in total internal reflection mode one can obtain useful quantitative response characteristics, as long as the polarization and phase gradient applied over the structure are precisely known.   We hence anticipate that we will be able to extend this method directly to quantify the magneto-electric scattering properties in Fourier space of  many interesting, but previously uncharted, plasmonic and metamaterial structures, such as SRRs, cut-wire pairs, structurally chiral objects, and oligomers where spectra are characterized by Fano resonances. We are currently extending the measurement technique  to telecommunication wavelengths where metamaterial structures have the  strongest electric and magnetic responses.

\addcontentsline{toc}{section}{Acknowledgments}
\ack
We thank Nir Rotenberg and Jean Cesario for fruitful discussions, and Jeroen Jacobs and Leon Huisman for their contributions to the building of the set up. This work is part of the research program
of the ``Stichting voor Fundamenteel Onderzoek der Materie
(FOM),'' which is financially supported by the ``Nederlandse
Organisatie voor Wetenschappelijk Onderzoek (NWO).'' AFK thanks NWO-Vidi and
STW/Nanoned.

\end{document}